\documentstyle[12pt]{article}
\input tcilatex
\begin{document}

\begin{center}
{\bf Comment about the ''Newtonian limit of String-Dilaton Gravity'' paper.}

\vspace*{1.5cm} S.M.KOZYREV

Scientific center gravity wave studies ''Dulkyn''.

e-mail: Kozyrev@e-centr.ru
\end{center}

\vspace*{1.5cm}

The objective of this brief comment is to point out several problems
associated with the general framework underpinning this paper\cite{1}. For
one, analyzing the static, spherically symmetric solution of scalar-tensor,
vector-metric as well as string-dilaton theories one can find the following.
When the energy-momentum tensor T=T$_\mu ^\mu $ vanishes the solutions of
field equations of these theories amongst the others equivalent to
Schwarzachild solution in general relativity. Moreover, for Jordan, Brans
Dicke scalar-tensor theory the Hawking theorem states that the Schwarzschild
metric is the only spherically symmetric solution of the vacuum field
equations. In this case no celestial-mechanical experiments to reveal a
difference between gravity theories is not presented possible, since all
Einsteins vacuum solution (Schwarzschild, Kerr, etc.) will satisfy these
theories too.

On the other hand inside the mater gravitation is greatly distinguished from
Einstein theory. Thus, in this case effects considered like the fifth force
and the differences between relativistic theories should be experimentally
tested in substance. This fact is matter of debate since in this point of
view the dark matter problem may be explain by variations of mater
properties inside the galaxies and galaxies clusters. For example, scalar
field, inside the matter has characteristics like gravitation permeability
of material similar electromagnetic permeability of material in Maxwell
theories of electromagnetism. Consequence there is a significant difference
in characteristics of galaxy models under similar boundary conditions for
various relativity theories.

Standard literature on gravitation describes any modification of celestial
mechanics due to relativistic theories is obtained by a perturbative
expansion valid for weak and quasi-stationary fields, known as the
post-Newtonian approximation. Solar-system experiments allowed one to map
out fairly completely weak-field gravity at the first post-Newtonian
approximation, i.e., to put stringent numerical constraints on a large class
of possible deviations from general relativity at order 1/c$^2$. Nordtvedt,
Will and others \cite{2} were led to provide rigorous underpinnings to the
operational significance of various theories, especially in solar system
context, developing the parameterized post Newtonian formalism as a
theoretical standard for expressing the predictions of relativistic
gravitational theories in terms which could be directly related to
experimental observations. However, for the case of equivalences vacuum
solutions we come to a conclusion that PPN parameters in empty space in
these relativistic theories are similar. Moreover, post Newtonian parameters
of the concrete gravitational theory under study will take different values
in external and internal areas of material objects.

Howsoever, it will be outlined here that the Newtonian limit cannot
reproduce all the solutions of the fully general relativistic theories. At
the same time it is easily proven that solution of the field equations for
spherically symmetric vacuum space around a point mass M for linearized
field equations (18) and (20) reduces not only to (22)-(24) but:

\begin{center}
\begin{eqnarray*}
h_{ik}=-\frac{2G_NM}{c^2}\frac{x_ix_k}{r^3}, \\
h_{00}=-\frac{2G_NM}{c^2}\frac 1r, \\
\psi =0, \\
h_{i0}=h_{0i}=0,i=k=1,2,3
\end{eqnarray*}
\end{center}

where r = $\sqrt{x_1^2+x_2^2+x_3^2}.$

This solution explicitly none depend on the parameters of gravity theory.
That is the case for instance of the Einsteins \cite{a1}$,$ scalar-tensor,
vector-metric theories too; spacetime of gravity theories in a weak field
approximation is similar in regions devoid of matter, and test particles do
not feel any distinctions of gravitational field. On the other hand for
Schwarzschild metric the field equations dont reduces to the Poisson
equation and in a presence of gravitational field Euclidean geometry
incorrect even in the first approximation. Hence, for Einsteins,
scalar-tensor, vector-metric as well as string-dilaton theories field
equations not reproduces Newtonian gravity when the low energy regimen is
consistently analyzed. However, weak field approximation of the
gravitational theories inside the matter do not reproduce General
Relativitys results but generalize those introducing corrections to the
Newtonian potential which could have interesting physical consequences. One
can find that the some theories give rise to terms capable of explaining the
flat rotation curve of galaxies with using a certain equation of state or
function of matter distributions even in weak field approximation. Using the
hypothesis that the not gravitational fields are the dark matter in galaxies
and solving the field equations for interior areas, we are able to reproduce
the rotation curves profile of stars going around galaxies.


\begin{thebibliography}{9}
\bibitem{1}  S.Capozziello, G. Lambiase, Int.J.Mod.Phys. D12 (2003) 843-852.

\bibitem{2}  Clifford M. Will, Theory and experiment in gravitational
physics, (2nd edn.), Cambridge University Press, Cambridge, 1993.

\bibitem{a1}  A.Einstein, Ann. D. Phys., 49, 769, 1916.
\end{thebibliography}
\end{document}